\documentclass{aa}  

\usepackage{graphicx}
\usepackage{txfonts}
\usepackage{amsmath}
\usepackage{gensymb}
\usepackage[normalem]{ulem}

\begin{document}

   \title{Dynamical friction from self-interacting dark matter}
    \titlerunning{Dynamical friction from SIDM}

   \author{Moritz S.\ Fischer
          \inst{\ref{inst:usm},\ref{inst:origins}} and
          Laura Sagunski\inst{\ref{inst:itp}}
          }
    \authorrunning{M.\ S.\ Fischer \& L.\ Sagunski}

    \institute{
        Universitäts-Sternwarte, Fakultät für Physik, Ludwig-Maximilians-Universität München, Scheinerstr.\ 1, D-81679 München, Germany\label{inst:usm}\\
        \email{mfischer@usm.lmu.de}
        \and
        Excellence Cluster ORIGINS, Boltzmannstrasse 2, D-85748 Garching, Germany\label{inst:origins}
        \and
        Institute for Theoretical Physics, Goethe University, 60438 Frankfurt am Main Germany\label{inst:itp}\\
        \email{sagunski@itp.uni-frankfurt.de}
    }

   \date{Received 28 June, 2024 / Accepted 15 July, 2024}

 
  \abstract
   {Merging compact objects such as binary black holes
   provide a promising probe for the physics of dark matter (DM). The gravitational waves emitted during inspiral potentially allow one to detect DM spikes around black holes. This is because the dynamical friction force experienced by the inspiralling black hole alters the orbital period and thus the gravitational wave signal.}
   {The dynamical friction arising from DM can potentially differ from the collisionless case when DM is subject to self-interactions. This paper aims to understand how self-interactions impact dynamical friction.}
   {To study the dynamical friction force, we use idealised $N$-body simulations, where we include self-interacting dark matter.}
   {We find that the dynamical friction force for inspiralling black holes would be typically enhanced by DM self-interactions compared to a collisionless medium (ignoring differences in the DM density). At lower velocities below the sound speed, we find that the dynamical friction force can be reduced by the presence of self-interactions.}
   {DM self-interactions have a significant effect on the dynamical friction for black hole mergers. Assuming the Chandrasekhar formula may underpredict the deceleration due to dynamical friction.}


   \maketitle
%

\section{Introduction}
Since the first direct detection of gravitational waves (GWs) from merging stellar-mass black holes (BHs) and neutron stars with the Laser Interferometer Gravitational-Wave Observatory \citep[LIGO;][]{Abbott_2016, Abbott_2017a, Abbott_2017b, Abbott_2019}, 100 years after the formulation of Einstein's theory of general relativity \citep{Einstein_1916} and about 40 years of the first indirect evidence \citep{Hulse_1975, Taylor_1976, Weisberg_1981}, GW astronomy has become an important and very promising field to probe fundamental physics.
The GWs emitted by binary mergers do not only allow for testing general relativity at unprecedented levels, but also to probe other physical processes. 
This in particular includes the physics of neutron stars. Binary mergers of those objects could allow probing the equation of state of ultrahigh-density matter \citep[e.g.][]{Baiotti_2022}.
In the future, GW detectors will become sensitive to sub-Hertz GW frequencies, thanks to space-based GW observatories such as the Laser Interferometer Space Antenna \citep[LISA;][]{Amaroseoane_2017} and Taiji \citep{Hu_2017}.
This will allow us to observe the evolution of intermediate- and extreme-mass ratio inspirals (I/EMRIs), in which a small stellar-mass compact object inspirals into a much larger intermediate/supermassive black hole, over long times and provide a sensitive probe of BH environments.

BHs might be `dressed' with DM, in other words, surrounded by a dense DM spike \citep[e.g.][]{Gondolo_1999, Bertone_2005, Bertone_2024b}.
The first robust evidence for an observation of a DM spike around a BH is claimed to be found by \cite{Chan_2024}, see also the work by \cite{Alachkar_2023}. They studied the supermassive black hole binary system OJ 287 with outburst observations covering a time span of 104 years.
The presence of a DM spike can impact the orbit of an inspiralling BH and make the GW signal sensitive to DM physics \citep[e.g.][]{Barausse_2014}.
This could in principle be detectable, for example for M87* \citep{Daghigh_2024}.
The inspiralling BH is decelerated by the dynamical friction force, which depends on the density and the velocity dispersion of the DM spike \citep{Chandrasekhar_1943}. In consequence, the orbital decay of an inspiralling BH allows probing properties of the DM spike and thus the particle nature of DM.

Potentially, such DM spikes could lead to a faster decay of the orbit at small separations ($\approx 1$ pc). This would provide an explanation for the final parsec problem, that is, how binary BHs can lose enough energy and angular momentum to reach separations small enough that the energy loss due to the emission of GWs is sufficient to make them merge within a sufficiently small timescale \citep[e.g.][]{Milosavljevi_2003}.
In this context, \cite{AlonsoAlvarez_2024} claim that collisionless DM is not sufficient to make the BH orbit decay fast enough because the DM spike would be destroyed by the friction energy injected from the BHs. Instead, to solve the final parsec problem, DM self-interactions would be needed to recreate the spike on a short enough timescale.

Self-interacting dark matter (SIDM) is a class of particle physics models assuming that DM has self-interactions beyond gravity. Initially, SIDM was proposed to solve problems on galactic scales, namely reducing the abundance of Milky Way satellites and the density in the centre of DM haloes \citep{Spergel_2000}. DM self-interactions have been studied in systems covering many orders of magnitude in mass, ranging from dwarf galaxies to galaxy clusters \citep[e.g.][]{Vogelsberger_2014, Fitts_2019, Robertson_2019, Zeng_2022, Ragagnin_2024, Sabarish_2024}.
Observations of those systems have been used to constrain the strength of the self-interactions \citep[e.g.][]{Sagunski_2021, Despali_2022, Kong_2024, Yang_2024S, Zhang_2024}.
For a review on SIDM, we refer the reader to \cite{Tulin_2018} and \cite{Adhikari_2022}.

Potentially, the range of systems used to constrain SIDM can be extended by merging binary BHs.
Based on pulsar timing array observations \citep{Agazie_2023, Reardon_2023, Antoniadis_2023}, it was claimed that a self-interaction cross-section of order $0.5$--$5.0 \, \mathrm{cm}^2 \, \mathrm{g}^{-1}$ would be needed to solve the final parsec problem \citep{AlonsoAlvarez_2024}. We note that this claim is made based on the assumption of the density spike being in the long-mean-free-path regime (see Sect.~\ref{sec:analytic_head_conduct}).
Interestingly, those binary systems could be a very sensitive probe for DM scattering, as we discuss in this paper.

When DM is self-interacting, the density profile of the DM spike may change compared to collisionless matter.
For SIDM, the slope of the density profile of the DM spike could be smaller than in the case of collisionless DM \citep{Shapiro_2014}.
The lower density would allow relaxing constraints from DM annihilation \citep{Alvarez_2021}.
Importantly, this also affects the dynamical friction, as it depends on the density of the DM spike, and thus also the GW signal \citep[e.g.][]{Macedo_2013, Yue_2018, Becker_2022}.
However, given the properties of the density spike, one may still need to calibrate the strength of the dynamical friction with simulations. \cite{Kavanagh_2020, Kavanagh_2024} run $N$-body simulations of collisionless DM to measure the Coulomb logarithm for a BH orbiting in a DM spike.

Several studies of dynamical friction for various DM models have been conducted. This includes ultralight DM \citep{Wang_2022, Boey_2024}, also with self-interactions \citep{Glennon_2024}, self-interacting scalar DM \citep{Boudon_2022, Boudon_2023b, Boudon_2023a, Kadota_2024}, super-fluid DM \citep{Berezhiani_2023}, and fuzzy DM \citep{Lancaster_2020}.
In this paper, we investigate the strength of the dynamical friction for a medium consisting of SIDM.

Often the binary BHs have been described in the Newtonian limit, however, depending on their evolution stage this may not be valid. Studies beyond Newtonian gravity have been undertaken by various authors \citep[e.g.][]{Traykova_2023, Montalvo_2024, Mukherjee_2024}.
For this study, we focus on the Newtonian regime, that is, when the separation of the two BHs is sufficiently large.
This allows us to build upon earlier work and simplifies the numerical description compared to the relativistic case. Extending our study to the relativistic case is left for future work.

The rest of this paper is structured as follows:
Sect.~\ref{sec:analytic_estimates} deals with the analytic description of dynamical friction and discusses it in the context of a BH moving through a DM spike.
In Sect.~\ref{sec:numerical_setup}, we describe the numerical setup that we use to study dynamical friction in the context of DM self-interactions. This includes the simulation code and the initial conditions (ICs) and parameters for the simulations. It follows the analysis of our simulations and the results (Sect.~\ref{sec:test_problems}).
In Sect.~\ref{sec:discussion}, we discuss the limitations of this work as well as
directions for further research.
Finally, we summarise and conclude in Sect.~\ref{sec:conclusion}.
Additional information is provided in the appendix.

\section{Analytic estimates} \label{sec:analytic_estimates}

In this section, we first describe the properties of DM spikes around BHs relevant to our study.
Secondly, we review analytic estimates of the dynamical friction force for collisionless and gaseous media.
Last, we discuss how to describe the strength of self-interactions and their impact on the dynamical friction arising from DM spikes acting on inspiralling BHs.

\subsection{DM spike}

\begin{figure}
    \centering
    \includegraphics[width=\columnwidth]{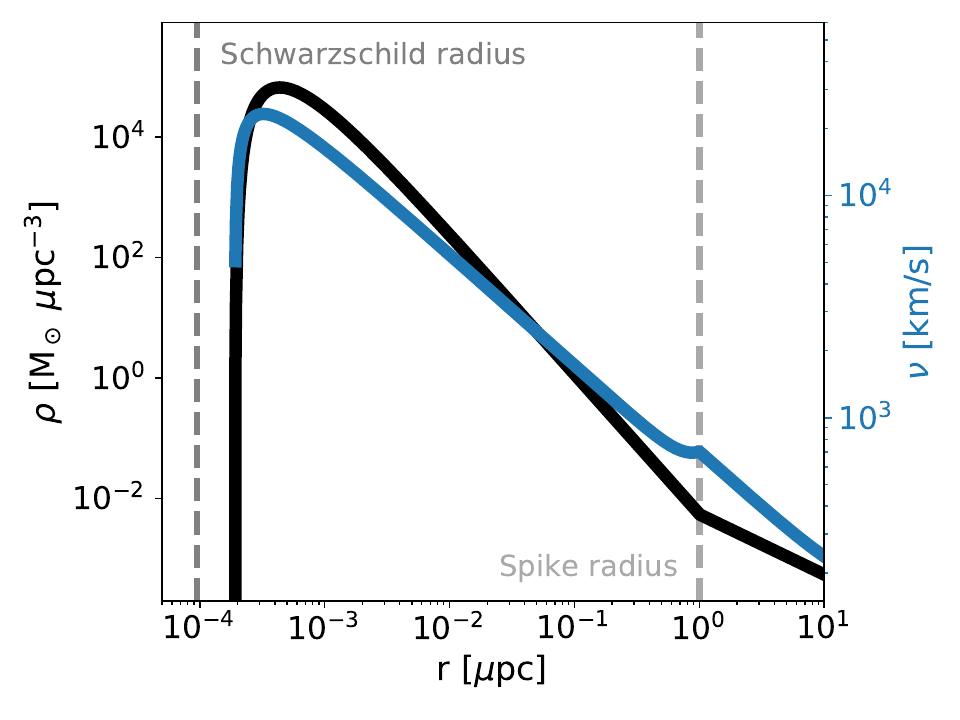}
    \caption{Density and one-dimensional velocity dispersion for a DM density spike as a function of radius according to Eq.~\eqref{eq:density_spike_rel}. We assume $r_\mathrm{sp} = 1 \mu \mathrm{pc}$, $\rho_\mathrm{sp} = 5.45 \times 10^{-3} \, \mathrm{M_\odot} \, \mu \mathrm{pc}^{-3}$, $\alpha_\mathrm{sp} = 7/3$, and $M_\mathrm{BH} = 10^3 \, \mathrm{M}_\odot$. In addition, we indicate the Schwarzschild radius $R_s = 2 \mathrm{G} M / c^2$.
    For computing the velocity dispersion, we followed \cite{Kremer_2022}. Here, we assumed an isotropic velocity distribution. In contrast to \cite{Kremer_2022}, we solved their Eq.~27 including higher-order terms instead of Eq.~28. Moreover, we show the results for $m / (k_\mathrm{B} T) \approx 6.53 \times 10^3 \, \mathrm{s}^2 \, \mathrm{km}^{-2}$, with $m$ being the rest mass of a physical DM particle, $k_\mathrm{B}$ the Boltzmann constant and $T$ the absolute temperature of the system.
    }
    \label{fig:density_spike}
\end{figure}

In the vicinity of a BH the DM density can potentially reach high values due to the gravitational influence of the BH. Such a DM spike is commonly parameterised by a power law,
\begin{equation} \label{eq:density_spike_nonrel}
    \rho_\mathrm{DM}(r) = \rho_6 \left(\frac{r}{r_6}\right)^{-\alpha_\mathrm{sp}}\, .
\end{equation}
The radius $r_6 = 10^{-6} \, \mathrm{pc}$ is a reference radius at which the density is given by $\rho_6$.
We note that with this choice, we follow \cite{Coogan_2022}.

When taking relativistic effects into account \citep{Sadeghian_2013} the spike profile becomes,
\begin{equation} \label{eq:density_spike_rel}
    \rho_\mathrm{DM}(r) = 
    \begin{cases}
        \, 0 & \textnormal{for} \quad r \leq 2 \, R_s \\
        \, \rho_{\rm sp} \left(1-\frac{2 R_s}{r}\right)^3 \left(\frac{r}{r_{\rm sp}}\right)^{-\alpha_\mathrm{sp}} & \textnormal{for} \quad   2 \, R_s < r \leq  r_{\rm sp}\\
        \, \frac{\rho_{s} \, r_{s}}{r} & \textnormal{for} \quad  r_{\rm sp} < r \ll r_s\\
    \end{cases}\,.
\end{equation}
For the relativistic corrections on the properties of the DM spike, we refer the reader to \cite{Eda_2015}, \cite{Tang_2021}, \cite{Capozziello_2023}, and \cite{John_2023}. In Fig.~\ref{fig:density_spike} we show the corresponding spike density and velocity dispersion as a function of radius.
We note that the density profile of the DM around the BH could be even more complicated when considering a realistic formation scenario of the BH \citep{Bertone_2024a}.

As we see in the next subsection the Mach number of the secondary BH moving through the density spike plays a crucial role in determining the dynamical friction force acting on it.
The Mach number, $\mathcal{M} = v_\mathrm{BH} / c_\mathrm{s}$, is given by the velocity, $v_\mathrm{BH}$, of the BH and the sound speed, $c_\mathrm{s}$.
The local speed of sound depends on the adiabatic index $\gamma$, and is given by
\begin{equation}
    c_\mathrm{s} = \sqrt{\gamma} \, \sigma_\mathrm{DM} \, .
\end{equation}
To estimate $c_\mathrm{s}$, we assume for simplicity that the local velocity distribution obeys a Maxwell-Boltzmann distribution with a one-dimensional velocity dispersion $\sigma_\mathrm{DM}$.
For the Mach number, $\mathcal{M}$, we make the assumption that the DM spike at sufficiently large radii is well described by Eq.~\ref{eq:density_spike_nonrel}.

Furthermore, we assume that the gravitational potential is dominated by the mass, $M_\mathrm{BH}$ of the primary (more massive) BH, and that the contribution of the DM spike and of the less massive inspiralling BH are negligible. Under these assumptions, the one-dimensional velocity dispersion of the DM is
\begin{equation}
    \sigma_\mathrm{DM} = \sqrt{\frac{\mathrm{G} \, M_\mathrm{BH}}{1+\alpha_\mathrm{sp}} \frac{1}{r}} \, .
\end{equation}
Assuming that the BH is moving on a circular orbit ($v_\mathrm{circ} = \sqrt{\mathrm{G} \, M(<r) /r}$), it follows that the Mach number is
\begin{equation} \label{eq:mach_number}
    \mathcal{M} = \sqrt{\frac{1+\alpha_\mathrm{sp}}{\gamma}} \,.
\end{equation}

The adiabatic growth model for a DM spike assuming collisionless DM suggests, for an NFW halo, a spike index of $\alpha_\mathrm{sp}=7/3$ \citep{Gondolo_1999, Fields_2014, Eda_2015}.
For SIDM, \cite{Shapiro_2014} derived the DM spike to be $\alpha_\mathrm{sp}=7/4$.
We note that this is based on the assumption that the self-interaction cross-section has the velocity dependence $\sigma \propto v^{-4}$.
Moreover, \cite{Shapiro_2014} assumed that the spike resides in the long-mean-free-path regime, in Sect.~\ref{sec:analytic_head_conduct}, we explain that this may eventually not always be the case.
This implies that the motion of the inspiralling BH is supersonic when being on a circular orbit (see Eq.~\ref{eq:mach_number}) given a spike index of $\alpha_\mathrm{sp}=7/3$ or $\alpha_\mathrm{sp}=7/4$.
Even for a smaller value of $\alpha_\mathrm{sp}$, it is not feasible to reach the subsonic regime on a circular orbit within the range typically considered for the spike index ($1 < \alpha_\mathrm{sp} < 3$).
However, a lower value for $\alpha_\mathrm{sp}$ might be possible depending on the formation scenario \citep{Ullio_2001}.

\subsection{Dynamical friction} \label{sec:analytic_estimates_dynfric}
The dynamical friction force for a perturber moving through a collisionless medium of constant density was derived by \cite{Chandrasekhar_1943}.
The strongly collisional case with an object moving through a gaseous medium was studied analytically by several authors \citep[e.g.][]{Rephaeli_1980, Just_1986, Just_1990}.
\cite{Ostriker_1999} derived an analytic description for this regime. Following her, the dynamical friction force can be expressed as
\begin{equation} \label{eq:dynfric}
    F_\mathrm{df} = - 4\pi\,\rho_\mathrm{DM} \left(\frac{\mathrm{G}\,m_\mathrm{BH}}{v_\mathrm{BH}}\right)^2 \, I \,.
\end{equation}
Here, $m_\mathrm{BH}$, is the mass of the perturber, in our case a BH. Its velocity is given by $v_\mathrm{BH}$ and the density of the perturbed medium, in our case DM, is denoted by $\rho_\mathrm{DM}$.
The factor $I$ depends besides other things such as the Mach number, $\mathcal{M}$, on the collisional nature of the perturbed medium and can be specified for the different cases.
In the collisionless case, $I$, is given by
\begin{multline} \label{eq:chandrasekhar}
    I_\mathrm{collisionless} = \ln{\left(\frac{r_\mathrm{max}}{r_\mathrm{min}}\right)} \left[ \mathrm{erf}(X) - \frac{2X}{\sqrt{\pi}} \, e^{-X^2}\right] \\ \textnormal{with} \quad X \equiv \frac{v_\mathrm{BH}}{\sigma_\mathrm{DM} \sqrt{2}} \,.
\end{multline}
The effective size of the BH is indicated by $r_\mathrm{min}$, and $r_\mathrm{max}$ denotes the effective size of the surrounding medium.
Where, $r_\mathrm{min}$, can be understood as the $90\degree$ deflection radius and $r_\mathrm{max}$ as the maximum impact parameter \citep{Binney_2008}.
We note that while Chandrasekhar's formula given by Eq.~\ref{eq:chandrasekhar} approximates the dynamical friction well, corrections to the Coulomb logarithm, $\ln(r_\mathrm{max}/r_\mathrm{min})$, can be necessary \citep{Just_2005, Just_2010}, for example for velocity distributions deviating from a Maxwell--Boltzmann distribution.
For corrections arising from fast moving DM particles on the Chandrasekhar formula, we refer the reader to the work by \cite{Dosopoulou_2023} on the dynamical friction in DM spikes.

For gaseous media, one distinguishes the subsonic case where the BH is moving with a velocity $v$ that is smaller than the speed of sound $c_\mathrm{s}$ and the supersonic case where it is faster than $c_\mathrm{s}$. In the first one, the Mach number, $\mathcal{M} = v / c_\mathrm{s}$ is smaller than unity and $I$ is given as
\begin{equation}
    I_\mathrm{subsonic} = \frac{1}{2} \ln{\left(\frac{1+\mathcal{M}}{1-\mathcal{M}}\right)} - \mathcal{M} \,.
\end{equation}
We note that this equation is based on the assumption that $(c_s - v_\mathrm{BH}) \, t$ exceeds the effective size, $r_\mathrm{min}$, of the perturber and that $(c_s + v_\mathrm{BH}) \, t$ is smaller than the effective size, $r_\mathrm{max}$, of the surrounding medium.
The time, $t$, indicates for how long the BH is perturbing the DM distribution.
In the supersonic case with $\mathcal{M} > 1$, we have
\begin{equation}
    I_\mathrm{supersonic} = \frac{1}{2} \, \ln{\left(1-\frac{1}{\mathcal{M}^2}\right)} + \ln{\left(\frac{v_\mathrm{BH}\,t}{r_\mathrm{min}}\right)}\,.
\end{equation}
This implicitly assumes $(v_\mathrm{BH} - c_s) \, t > r_\mathrm{min}$ and $(v_\mathrm{BH} + c_s) \, t < r_\mathrm{max}$.

\begin{figure}
    \centering
    \includegraphics[width=\columnwidth]{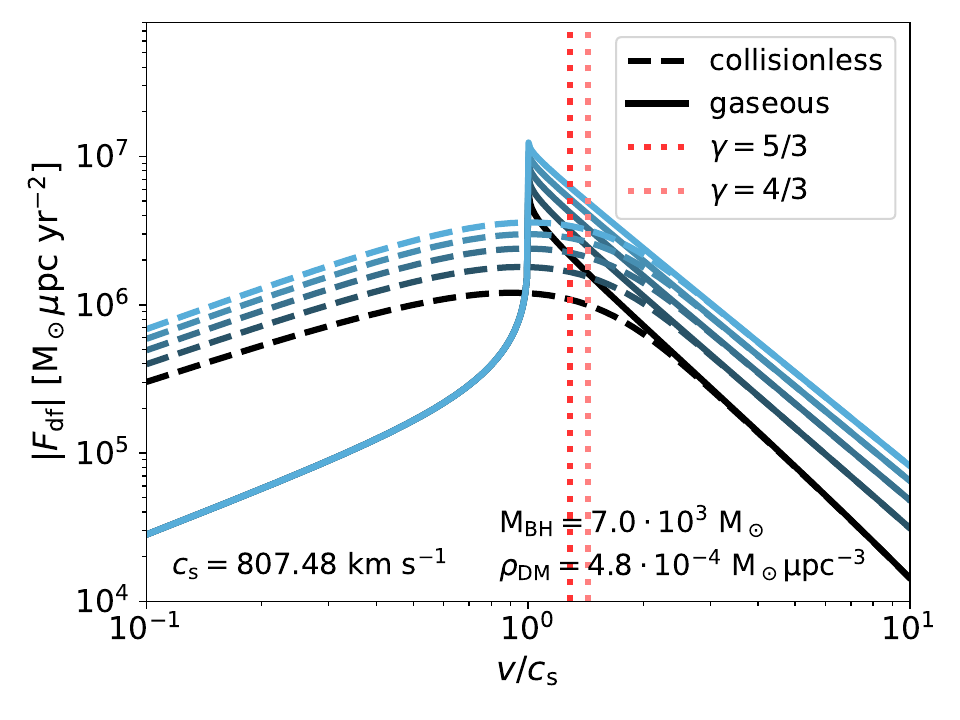}
    \caption{Dynamical friction force as a function of the Mach number, i.e.\ the velocity in units of the sound speed. The dashed lines indicated the dynamical friction for a collisionless medium according to the Chandrasekhar formula \citep{Chandrasekhar_1943} and the solid lines are for a gaseous medium as given by \cite{Ostriker_1999}. We note that for the plot, we assume $r_\mathrm{max} = \mathcal{M} \, c_s \, t$ and $\ln(r_\mathrm{max} / r_\mathrm{min}) = \ln(\mathcal{M}) + C$. The line colours correspond to different values of $C \in \{4.0,6.0,8.0,10.0,12.0\}$. In addition, the red dotted lines indicate the velocity of a secondary BH moving on a circular orbit for different adiabatic indices.}
    \label{fig:ostriker}
\end{figure}

In Fig.~\ref{fig:ostriker}, we plot the drag force computed according to \cite{Ostriker_1999} for our test setup used in Sect.~\ref{sec:test_problems}.
We can see that the dynamical friction for velocities slightly above the sound speed is becoming fairly strong in the gaseous case and sharply drops when the velocity falls below the speed of sound.

In the previous subsection, we have seen that the BH would typically move with a supersonic velocity.
In consequence, the inspiralling BH experiences enhanced dynamical friction if the DM is not collisionless. However, if the BH is on an eccentric orbit, it might partially, when it is close to its apocentre, move with a subsonic velocity.
We also note that dynamical friction can contribute to a circularisation of the orbit \citep{Becker_2022, Karydas_2024}.

In general, the dynamical friction acting on inspiralling BHs may not only arise from the DM spike but also from the accretion disk. The latter can also be described with the Ostriker formula \citep[e.g.][]{Szolgyen_2022, Becker_2023}.

\subsection{The strength of the self-interactions} \label{sec:analytic_head_conduct}
A remaining question is how close a given cross-section would be to the collisionless or gaseous regime.
This means we need to understand how much the trajectory of the DM particles is impacted by gravity compared to the scattering. A common measure for this in the context of SIDM haloes is the Knudsen number \citep[e.g.][]{Koda_2011}.
It is given by
\begin{equation} \label{eq:knudsen_number}
    \mathrm{Kn} = \sqrt{\frac{4 \pi \mathrm{G}}{\rho \sigma_\mathrm{DM}^2}} \left(\frac{\sigma_\mathrm{eff}}{m}\right)^{-1} \,.
\end{equation}
This is comparing the ratio of the mean free path to the Jeans length. The latter is a measure for how susceptible the DM medium is to gravitational collapse.
We note that the Knudsen number, Kn, is independent of the mass of the perturber, i.e.\ the BH mass.
However, the strength of the dynamical friction depends on it. 
For Eq.~\ref{eq:knudsen_number} we used the effective cross-section per mass,  $\sigma_\mathrm{eff}/m$, for defining the Knudsen number.
The effective cross-section \citep{Yang_2022D} allows one to map cross-sections with different velocity dependencies onto each other and is given by 
\begin{equation} \label{eq:sigma_eff}
    \sigma_\mathrm{eff} = \frac{\langle v^5 \sigma_\mathrm{V}(v)\rangle}{\langle v^5 \rangle} \, .
\end{equation}
Where an average integrating over the distribution of relative velocities, $v$, and weighting the cross-section with $v^5$ is computed assuming that the relative velocities follow a Maxwell--Boltzmann distribution. Furthermore, $\sigma_\mathrm{V}$ is the viscosity cross-section,
\begin{equation} \label{eq:sigma_v}
    \sigma_\mathrm{V} = 3\pi \int_{-1}^1 \frac{\mathrm{d}\sigma}{\mathrm{d}\Omega} \sin^2 \theta \, \mathrm{d}\cos\theta \,.
\end{equation}
Here, we follow the normalisation of \cite{Yang_2022D} implying that viscosity and total cross-section have equal values for isotropic scattering ($\left.\sigma_\mathrm{V}\right|_{\rm iso} =\left.\sigma_\mathrm{tot}\right|_{\rm iso}$).
The viscosity cross-section has the advantage that it often provides a good matching between different angular dependencies \citep[e.g.][]{Yang_2022D, Colquhoun_2020} and is thus a good characteristic for the strength of the scattering.
If the Knudsen number is large, i.e.\ $\mathrm{Kn} \gg 1$, a system is considered to be in the long-mean-free-path regime, i.e.\ gravity dominates over the scattering. In contrast, if $\mathrm{Kn} \ll 1$, the system would be in the short-mean-free-path regime and scattering dominates over gravity.
So we expect that the system is close to the collisionless case when the Knudsen number is larger, and for small numbers falls into the gaseous regime.

For the DM spike, the Knudsen number becomes smaller when getting closer to the primary BH as density and velocity dispersion increase.
The Knudsen number scales as $\mathrm{Kn} \propto r^{(1+\alpha_\mathrm{sp})/2}$ when assuming a density spike profile as given by Eq.~\eqref{eq:density_spike_nonrel} and the gravitational potential being dominated by the BH.
This implies that the behaviour of the dynamical friction could transition from the collisionless regime to the gaseous one while the secondary BH is inspiralling. However, it depends on the strength of the DM self-interactions.

Last, we note that also in the case of a primary BH hosting a DM spike, the heat conduction due to self-interactions being in the long-mean-free-path or short-mean-free-path regime is independent of the mass of the primary BH if Eq.~\ref{eq:knudsen_number} holds.
In other words, the Knudsen number does not depend on the mass of the primary BH. We note that this is in contradiction to the assumption by \cite{Shapiro_2014} and \cite{Shapiro_2018}. Instead of using the Jeans length for the Knudsen number, they approximated the gravitational scale height as the distance, $r$, to the primary BH. In fact, their Knudsen number is given as the mean free path divided by $r$. The influence of the BH changing the gravitational scale height can be motivated by studies of star clusters \citep[e.g.][]{Bettwieser_1986}. It is unclear if this is in general the same for the SIDM case.
If instead Eq.~\ref{eq:knudsen_number} holds, a DM spike given a cross-section within the typically studied range of $\approx$ 0.1--100 cm$^2$g$^{-1}$ \citep[e.g.][]{Tulin_2018, Adhikari_2022} could be to a large extent rather in the short-mean-free-path regime than in the long-mean-free-path regime.
This would be in line with the work by \cite{Pollack_2015}, where they studied a DM model with a fraction of the matter being ultra-strongly self-interacting. Their heat conductivity does only depend on the density of the ultra-strongly interacting matter component.

\section{Numerical setup} \label{sec:numerical_setup}
In this section, we first describe our simulation code with its implementation of DM self-interactions. Next, it follows the setup for our simulations to understand the role of self-interactions for dynamical friction. 

\subsection{Simulation code}

We used the cosmological $N$-body code \textsc{OpenGadget3} \citep[see][and the references therein]{Groth_2023}.
The SIDM module of the code has been described by \citet{Fischer_2021a, Fischer_2021b, Fischer_2022, Fischer_2024a}.
We employed it to model DM interactions beyond gravity.
The code can simulate both large and small-angle scatterings. To the first one, we also refer to as rare scattering since a single scattering event on average leads to a larger energy and momentum transfer between the particles compared to small-angle scattering, which must be much more frequent to have a similar effect. For the first one, the angular dependence is modelled explicitly by drawing random angles with respect to the differential cross-section, for the second one an effective description is used based on a drag-like behaviour \citep[see also][]{Kahlhoefer_2014}.
The interactions between the numerical particles are modelled on a pair-wise basis. We have computed the scattering probability for the rarely self-interacting dark matter (rSIDM) and the drag force for the frequently self-interacting dark matter (fSIDM) using a spline kernel \citep{Monaghan_1985}. Its size is determined with the next neighbour approach using $N_\mathrm{ngb} = 64$ neighbours.
This choice is known to provide accurate results \citep{Fischer_2021a}.
We employed a separate time step criterion for SIDM to keep the interaction probability per interaction sufficiently low.
The SIDM module is designed to explicitly conserve linear momentum and energy, even when executed in parallel.

To model the gaseous regime when self-interactions are very strong, we used smoothed-particle hydrodynamics \citep[SPH,][]{Gingold_1977}. Compared to \textsc{gadget-2} \citep{Springel_2005}, an updated SPH formulation \citep{Beck_2016} with an updated viscosity scheme \citep{Dolag_2005, Beck_2016} was employed. We chose a Wendland $C^6$ kernel \citep{Dehnen_2012} with its size being determined by the, $N_\mathrm{ngb} = 230$, next neighbours.

\subsection{Simulation setup}

To test the dynamical friction, we simulated a BH travelling through a constant density consisting of DM.
The mass density is set to $\rho = 4.815 \times 10^{-4} \, \mathrm{M_\odot} \, {\mu\mathrm{pc}}^{-3}$.
The velocities of the DM particles follow a Maxwell-Boltzmann distribution, and their one-dimensional velocity dispersion is chosen to be $\sigma_\mathrm{1D} = 625.5 \, \mathrm{km} \, \mathrm{s}^{-1}$.
The BH is represented by a single particle with a mass of $m_\mathrm{BH} = 7 \times 10^3 \, \mathrm{M_\odot}$. Initially, it is moving with a velocity of $v_\mathrm{ini} = 1037 \, \mathrm{km} \, \mathrm{s}^{-1}$.
These values are selected because they lead to a strong dynamical friction force, and we expect to find a significant impact in our simulations.
Furthermore, we chose a cubic simulation volume with a side length of $l_\mathrm{box} = 829.5 \, \mu\mathrm{pc}$ for most of our simulations but also used a box length of $l_\mathrm{box} = 663.6 \,\mu\mathrm{pc}$.
We varied the box length to study its impact on the dynamical friction force.
In addition, periodic boundary conditions were employed.
We note that this also implies a periodic computation of the gravitational forces. Otherwise, the distribution of DM would easily collapse.
The DM is resolved by $N_\mathrm{DM} = 5 \times 10^6$ particles, which leads to a numerical particle mass of $m_\mathrm{DM} = 5.5 \times 10^{-2} \, \mathrm{M_\odot}$. For testing the accuracy of our simulations, we also performed runs with $N_\mathrm{DM} = 1 \times 10^6$ DM particles, i.e.\ a five times higher particle mass.
We set the Plummer-equivalent gravitational softening length to $\epsilon = 24 \, \mu\mathrm{pc}$. 
The gravitational forces are computed between all particles, i.e.\ we also take the self-gravity of the perturbed medium into account.
In principle, the self-gravity can be relevant for the resulting dynamical friction \citep[see][]{Just_1990}. However, we do not separately study its contribution.

We note that the $N$-body representation of the constant density is subject to fluctuation due to the sampling noise. The fluctuations can collapse under gravitational influence. We have checked that, for our choice of the box length and resolution, this does not occur within the time we simulate. 

We ran simulations with a very anisotropic cross-section by employing the numerical scheme for frequent self-interactions developed by \cite{Fischer_2021a}. Here the strength of the self-interactions is specified in terms of the momentum transfer cross-section,
\begin{equation} \label{eq:momentum_transfer_cross_section}
\sigma_\mathrm{T}=2 \pi \int_{-1}^{1} \frac{\mathrm{d} \sigma}{\mathrm{d} \Omega_{\mathrm{cms}}}\left(1-|\cos \theta_{\mathrm{cms}}| \right) \mathrm{d} \cos \theta_{\mathrm{cms}}
\, .
\end{equation}
We simulated velocity-independent elastic scattering with cross-sections of $\sigma_\mathrm{V} / m \in \{3 \times 10^{-9}, 3 \times 10^{-8}, 3 \times 10^{-7}, 9 \times 10^{-7}, 3 \times 10^{-6}\} \, (\mathrm{cm}^2 \, \mathrm{g}^{-1})$.\footnote{To make it easier to compare cross-sections of different angular dependencies, we do not specify the cross-section in terms of the momentum transfer cross-section but use the viscosity cross-section (see Eq.~\ref{eq:sigma_v}).} In addition, we simulated velocity-independent isotropic cross-sections with a total cross-section of $\sigma_\mathrm{V} / m \in \{2 \times 10^{-8}, 3 \times 10^{-8}, 2 \times 10^{-7}, 3 \times 10^{-7}\} \, (\mathrm{cm}^2 \, \mathrm{g}^{-1})$ to investigate the role of the angular dependence. We summarise all the cross-sections used in Table~\ref{tab:cross-sections}.

In addition, we also investigated velocity-dependent self-interactions. Here, we used the implementation by \cite{Fischer_2024a}.
We parameterise the velocity dependence of the viscosity cross-section as
\begin{equation} \label{eq:veldep}
    \frac{\sigma_\mathrm{V}}{m} = \frac{\sigma_0}{m} \left(1+ \left( \frac{v}{w} \right)^2 \right)^{-2} \,.
\end{equation}
For the velocity parameter, $w$, we choose $w \in \{300, 461.1, 683.2, 1483.5\} \, (\mathrm{km} \, \mathrm{s}^{-1})$.
We note that the mean relative velocity of the DM particles in the ICs is $\langle v \rangle = 1411.6 \, \mathrm{km} \, \mathrm{s}^{-1}$. For all simulations with a velocity-dependent cross-section, we chose the value of $\sigma_0$ to result in an effective cross-section \citep[see][]{Yang_2022D} of $\sigma_\mathrm{eff} / m = 3 \times 10^{-8} \, \mathrm{cm}^2 \, \mathrm{g}^{-1}$.
A summary of all cross-sections is given in Table~\ref{tab:cross-sections}.

We note that all the cross-sections mentioned above are many orders of magnitude smaller than the tightest constraints on SIDM in the current literature \citep[e.g.][]{Harvey_2019, Andrade_2021, Sagunski_2021, Eckert_2022, Gopika_2023}.
Cross-sections as large as the ones typically discussed in the astrophysical context imply a mean free path substantially shorter than the kernel size, and may no longer be faithfully simulated with the standard SIDM schemes \citep[see our Appendix~\ref{sec:mean_free_path} as well as][]{Fischer_2024b}.
In addition, the simulations become more costly with increasing cross-section as the required time-step decreases. To model the regime of such a large cross-section, we use SPH. For the ICs, we set the initial velocity of the SPH particles to zero and chose their internal energy to correspond to the same velocity dispersion as in the SIDM setup.

\begin{table}
    \caption{Simulated cross-sections.}
    \label{tab:cross-sections}
    \centering
    \begin{tabular}{ccc}
        \hline\hline
        Type & $\sigma_\mathrm{V}/m$ & $w$ \\
        & [cm$^2$g$^{-1}$] & [km s$^{-1}$] \\ \hline
        Collisionless & 0.0 & -- \\
        Frequent & $3 \times 10^{-9}$ & -- \\
        Rare & $2 \times 10^{-8}$ & -- \\
        Frequent & $3 \times 10^{-8}$ & -- \\
        Rare & $3 \times 10^{-8}$ & -- \\
        Rare & $2 \times 10^{-7}$ & -- \\
        Frequent & $3 \times 10^{-7}$ & -- \\
        Rare & $3 \times 10^{-7}$ & -- \\
        Frequent & $9 \times 10^{-7}$ & -- \\
        Frequent & $3 \times 10^{-6}$ & -- \\
        Frequent & $6 \times 10^{-5}$ & 300 \\
        Rare & $6 \times 10^{-5}$ & 300 \\
        Frequent & $1.2 \times 10^{-5}$ & 461.1 \\
        Rare & $1.2 \times 10^{-5}$ & 461.1 \\
        Frequent & $3 \times 10^{-6}$ & 683.2 \\
        Rare & $3 \times 10^{-6}$ & 683.2 \\
        Frequent & $3 \times 10^{-7}$ & 1483.5 \\
        Rare & $3 \times 10^{-7}$ & 1483.5 \\
        Fluid & -- & -- \\
        \hline
    \end{tabular}
    \tablefoot{The table gives the different cross-sections we used for the simulations of our test setup. The first column specifies the type of scattering, where frequent refers to a very anisotropic cross-section and rare to an isotropic one. The second column gives the viscosity cross-section (Eq.~\ref{eq:sigma_v}). In the case of the velocity-dependent cross-sections, the second column corresponds to $\sigma_0 / m$ of Eq.~\eqref{eq:veldep}. The last column gives the parameter $w$ of Eq.~\eqref{eq:veldep} for the velocity-dependent cross-sections.}
\end{table}

\section{Test problems} \label{sec:test_problems}

In this section, we present the results from our simulations following the setup described in Sect.~\ref{sec:numerical_setup}. We begin with the collisionless case and investigate the impact of variations in the numerical setup. Subsequently, we study the role of the strength of the self-interactions and its impact on the density wake. Furthermore, we investigate the matching of different angular and velocity dependencies onto each other in the case of dynamical friction.

\subsection{The collisionless case}

\begin{figure}
    \centering
    \includegraphics[width=\columnwidth]{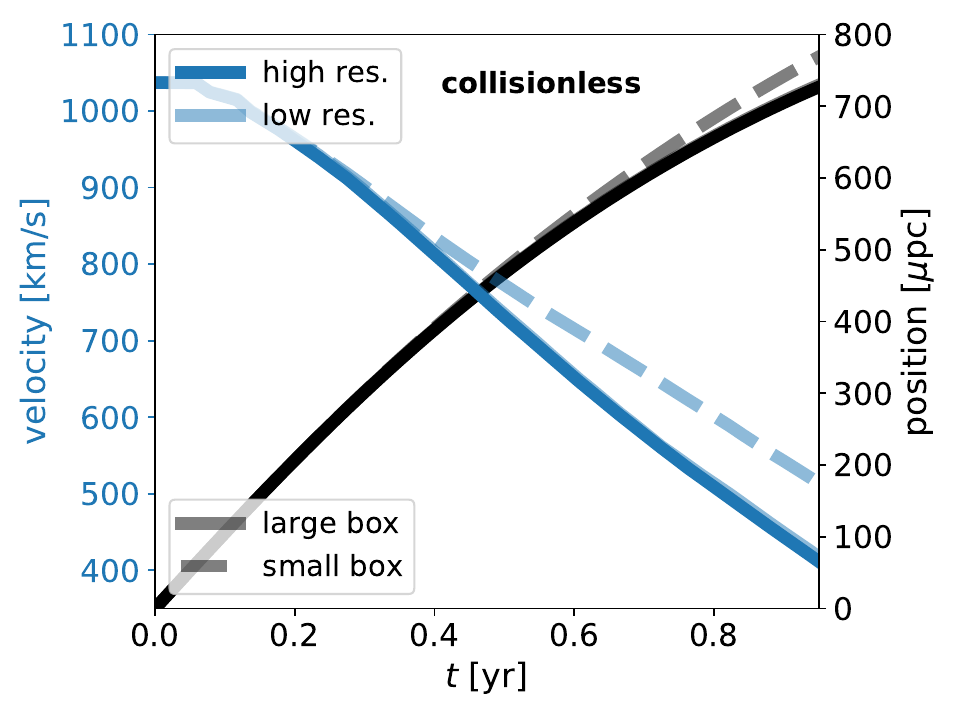}
    \caption{Velocity and position of the BH in our test setup as a function of time. All simulations are for collisionless DM. The solid lines are from the setup with a box length of $l=829.5 \mu\mathrm{pc}$. In contrast, the dashed lines correspond to a setup with a smaller box size of  $l=663.6 \mu\mathrm{pc}$. The high-resolution results correspond to $N_\mathrm{DM}= 5 \times 10^6$ DM particles, and the low-resolution ones to $N_\mathrm{DM} = 10^6$. We note that in the smaller box setup, the low resolution implies a higher mass resolution than in the large box setup.}
    \label{fig:cdm_vel_pos}
\end{figure}

In our test setup as described in Sect.~\ref{sec:numerical_setup}, the BH is travelling through an initially constant density.
The BH perturbs the DM and creates a density wake, leading effectively to a force decelerating the BH.
This dynamical friction force depends on the velocity of the BH relative to the velocity dispersion of the perturbed DM density (see Fig.~\ref{fig:ostriker}).
In our setup, we first expect the dynamical friction to become stronger while the BH is slowing down.
After the BH velocity has decreased enough, it enters the phase where the dynamical friction becomes weaker with decreasing velocity.

In Fig.~\ref{fig:cdm_vel_pos}, we show the position and velocity of the BH in our test setup for collisionless matter. As expected, the velocity of the BH decreases over the course of the simulation. The results for two different box sizes are shown. We find that the dynamical friction is reduced for a smaller box size. This is not surprising, as $r_\mathrm{max}$ in the Coulomb logarithm of the Chandrasekhar formula, Eq.~\eqref{eq:chandrasekhar}, depends for our setup effectively on the box size because the box size limits the maximal possible impact parameter.
The results agree qualitatively with the Chandrasekhar formula, after an initial phase during which the density wake forms, implying that Eq.~\eqref{eq:chandrasekhar} provides a good description when the Coulomb logarithm is taken as a free parameter.
For the large box, we show the results using two different resolutions. In both cases, the trajectory of the BH is very small, such that we can safely assume that the resolution for our study is large enough.

\begin{figure}
    \centering
    \includegraphics[width=\columnwidth]{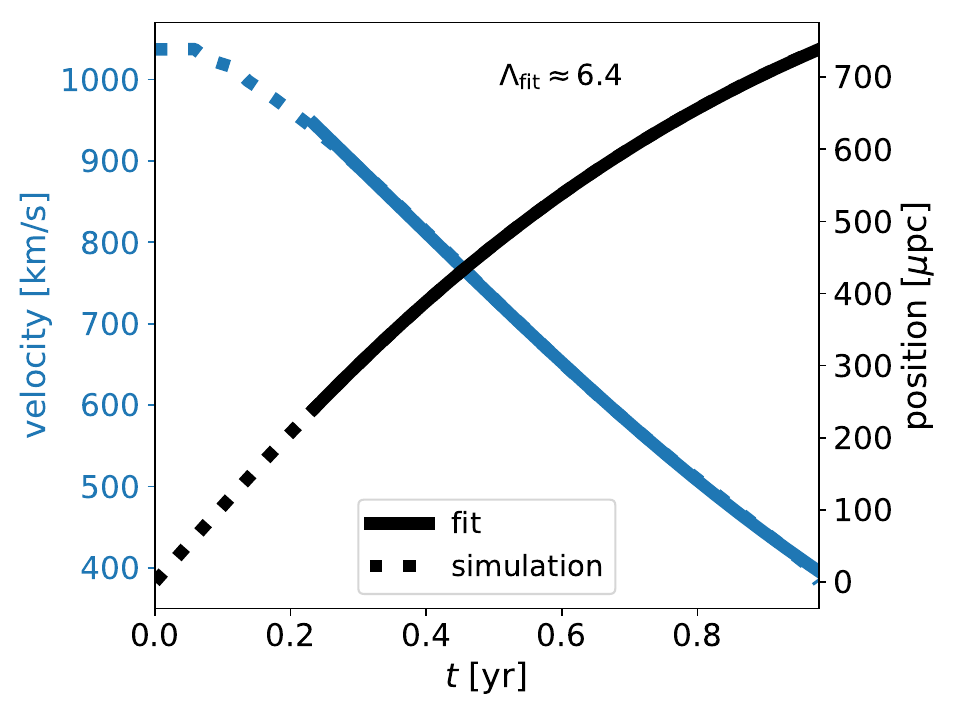}
    \caption{As in Fig.~\ref{fig:cdm_vel_pos}, position and velocity of the BH as a function of time. The high-resolution setup with the large box size of $l=829.5 \mu\mathrm{pc}$ is displayed. We fitted Chandrasekhar's formula (Eq.~\ref{eq:dynfric} and \ref{eq:chandrasekhar}) to the simulation results to determine the Coulomb logarithm, $\ln(\Lambda)$, with $\Lambda = r_\mathrm{max}/ r_\mathrm{min}$. Besides the simulation result (dotted lines), we also show the fitted curves (solid lines).}
    \label{fig:cdm_fit}
\end{figure}

We also quantitatively determined the Coulomb logarithm, $\ln(\Lambda) = \ln(r_\mathrm{max}/ r_\mathrm{min})$, by fitting $\Lambda$ to the simulation data for the run with the high resolution ($N_\mathrm{DM} = 5 \times 10^6$) and the large box size ($l=829.5 \mu\mathrm{pc}$). Here we did not use all of the simulation data but exclude the first phase of the simulation where the density wake forms and used the BH velocities for the fit only.
The simulation results together with the fitted curves are displayed in Fig.~\ref{fig:cdm_fit}. Our fit yielded a value of $\Lambda_\mathrm{fit} \approx 6.4$.
Besides we can analytically estimate the expected value for $\Lambda$. It can be expressed as \citep{Binney_2008}
\begin{equation}
    \Lambda = \frac{b_\mathrm{max} \, v^2_\mathrm{typ}}{\mathrm{G} \, M}.
\end{equation}
We approximated the maximum impact parameter, $b_\mathrm{max}$, with half the size of the box length, $l$. The typical velocity, $v_\mathrm{typ}$, between the BH and the DM particles is approximately $\approx 10^3 \, \mathrm{km} \, \mathrm{s}^{-1}$. The mass, $M$, of the perturber might effectively be larger than the BH mass as DM particles are bound to the BH. We discuss this in Sect.~\ref{sec:results_wake_properties}. Here we can assume, $M\approx 2 \times m_\mathrm{BH}$. Last, $\mathrm{G}$, denotes the gravitational constant. Based on these assumptions we obtained a value of $\Lambda \approx 6.9$, which agrees well with our simulation result ($\Lambda_\mathrm{fit} \approx 6.4$), especially given that the logarithm of that value matters for the dynamical friction.

\subsection{Dependence on the strength of the self-interactions}

\begin{figure}
    \centering
    \includegraphics[width=\columnwidth]{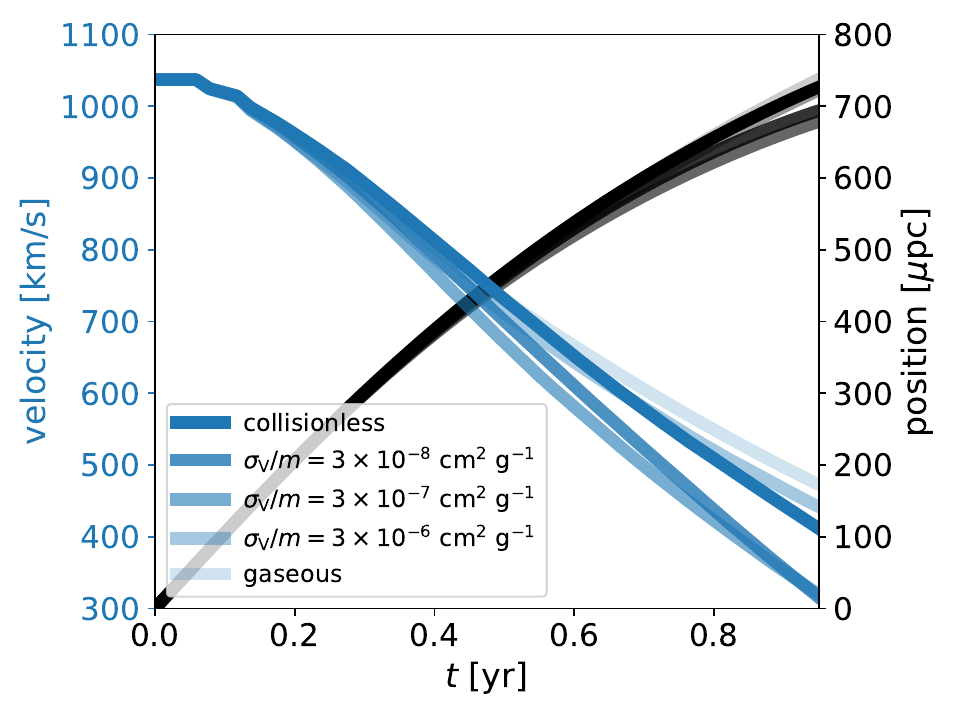}
    \caption{Similar to Fig.~\ref{fig:cdm_vel_pos}, position and velocity of the BH particle as a function of time. The results are shown for several interaction strengths of the DM, indicated by the opacity of the lines. All simulations are for the large box with the high resolution of $N_\mathrm{DM} = 5 \times 10^6$ particles. For the simulation of gaseous matter, we used the same number of SPH particles.}
    \label{fig:all_vel_pos}
\end{figure}

Given that the dynamical friction force differs between the collisionless and the gaseous case as described in Sect.~\ref{sec:analytic_estimates}, we expect that self-interactions can have an impact on the strength of the dynamical friction force.
To study this in detail, we run simulations with SIDM.
In Fig.~\ref{fig:all_vel_pos}, we show the results for fSIDM with different values for the momentum transfer cross-section.
We also include the collisionless and gaseous case.
It is visible that the self-interactions can enhance or reduce the dynamical friction, depending on the strength of the cross-section.
For the gaseous regime, we expect the dynamical friction to be enhanced above the sound speed and reduced below.
Accordingly, we find a larger velocity of the BH at the end of the simulation, when the BH velocity has fallen well below the sound speed of the DM.

\begin{figure*}
    \centering
    \includegraphics[width=\textwidth]{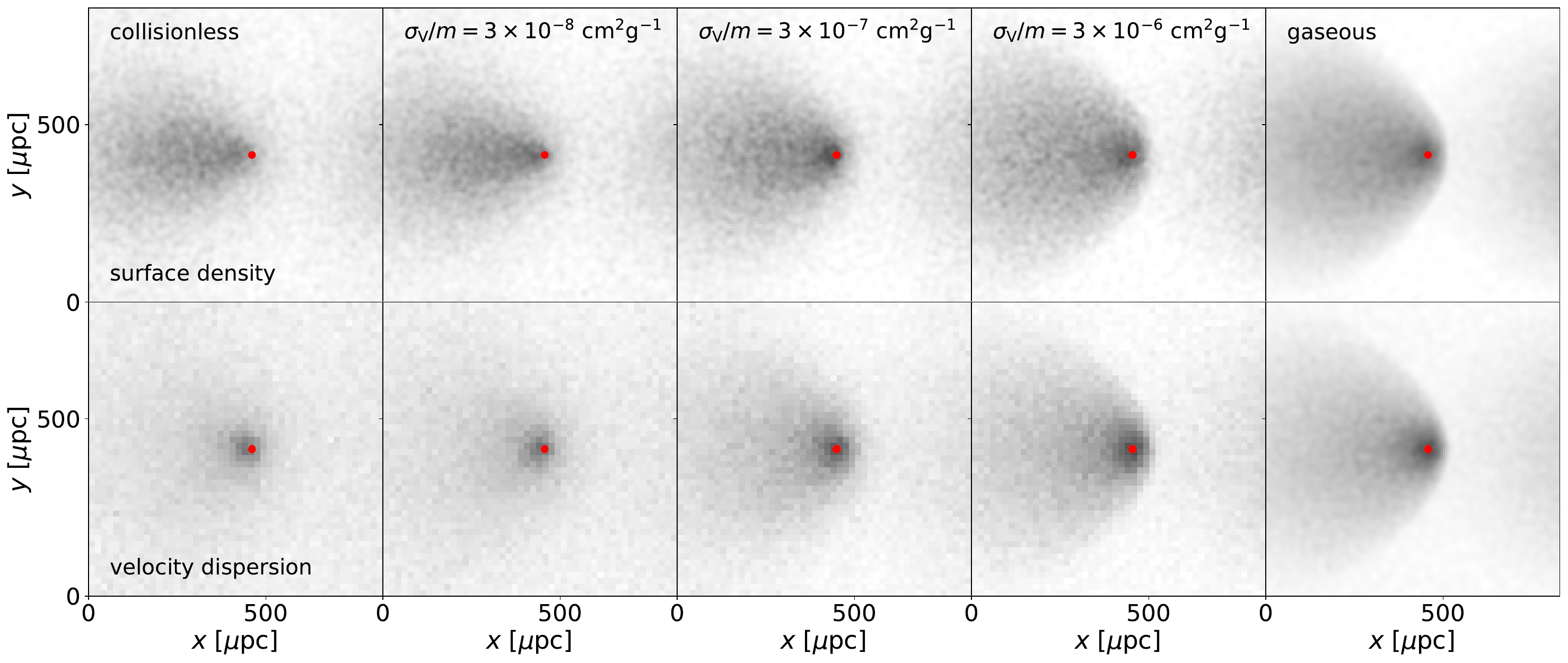}
    \caption{Surface density (top row) and the velocity dispersion (bottom row) for our test simulations at a time of $t=0.49\,\mathrm{yr}$. The panels display the different types of matter we investigate, with increasing collisionality from the left to the right. The density and one dimensional velocity dispersion are indicated following a logarithmic colour scale spanning a factor of two. For all panels of the same quantity, the same colour scale is used. In addition, we indicate the position of the BH particle with a red dot. We provide the time evolution of the surface density as a video available online.}
    \label{fig:sdens_map}
\end{figure*}

A detailed investigation of the dynamic friction force follows. But before, we take a look at the DM density perturbed by the BH.
For the same simulations as shown in Fig.~\ref{fig:all_vel_pos}, we show the surface density of the DM in the top row of Fig.~\ref{fig:sdens_map}.
Here, we can see that the shape and density of the wake depend on the strength of the self-interactions.
We find that for $\sigma_\mathrm{V}/m \in \{3 \times 10^{-8}, 3 \times 10^{-7}\} \, (\mathrm{cm}^2 \, \mathrm{g}^{-1})$ the densities in the wake are higher and at the same time the deceleration is stronger.
For these small cross-sections, the scattering leads to more low-velocity particles in the density wake compared to CDM effectively enhancing the deceleration of the BH.

In contrast, for the cases of even stronger self-interactions, we find that the densities in the wake are decreasing and the deceleration of the BH is weaker.
In principle, this can be seen from Fig.~\ref{fig:sdens_map}, but we study this quantitatively in Sect.~\ref{sec:results_wake_properties}.
We also note that the shape of the density wake changes for an increasing self-interaction strength, impacting the dynamical friction force as well.
The reduced dynamical friction for the strong cross-sections is in line with the expectation of a weaker deceleration in the gaseous case for the subsonic regime.

In addition to the surface density, we show the velocity dispersion in the bottom row of Fig.~\ref{fig:sdens_map}.
To compute the velocity dispersion, we first subtracted the bulk motion.
In the gaseous case, we computed the velocity dispersion directly from the temperature of the SPH particles.
Similar to the density plots, it is visible how the wake changes when self-interactions are present.
For example, the velocity dispersion in the vicinity of the perturber increases with cross-section and the shape of the wake changes.

\begin{figure}
    \centering
    \includegraphics[width=\columnwidth]{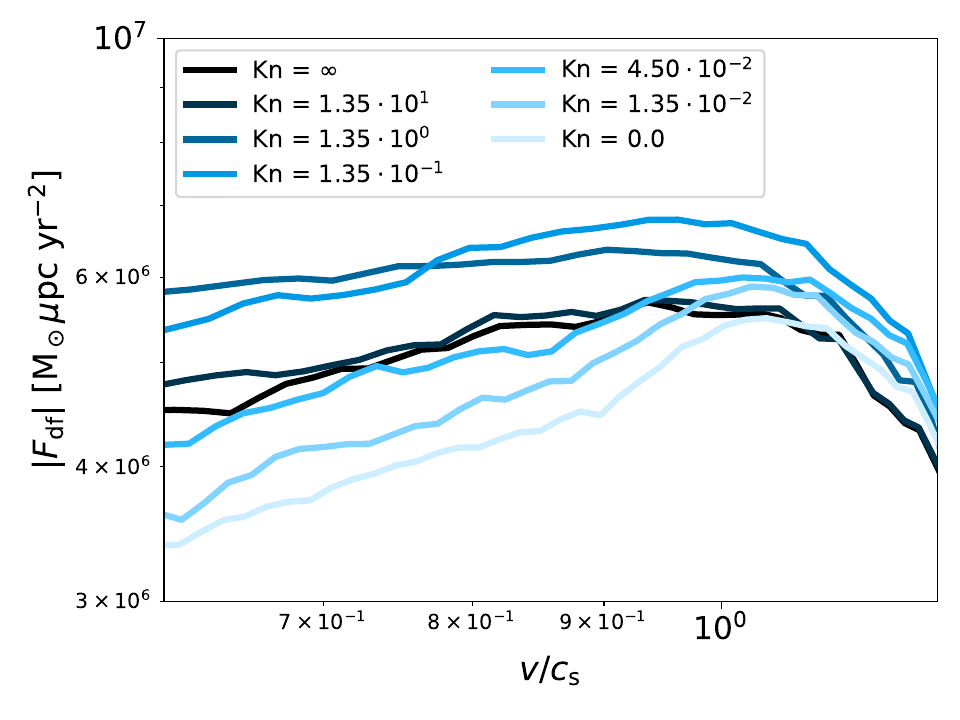}
    \caption{Dynamical friction force as a function of the Mach number for our simulations. This plot is analogue to Fig.~\ref{fig:ostriker}, which gives the analytic description of the dynamical friction force. Different cross-sections are shown, indicated by the Knudsen number. Here, $\mathrm{Kn} = \infty$ corresponds to a collisionless medium and $\mathrm{Kn} = 0.0$ to a gaseous medium. We note that in the simulation the BH initially starts at a high velocity, i.e.\ at the right side of the plot and slows down throughout the simulation, i.e.\ moves towards the left side of the plot.
    }
    \label{fig:dforce_vel}
\end{figure}

For a more detailed analysis of the dynamical friction force and its dependence on the cross-section, we computed the force and show it as a function of the velocity (Fig.~\ref{fig:dforce_vel}). The force is computed from the momentum of the BH and its change between consecutive snapshots.
We note that the strongest SIDM cross-section shown is $\sigma_\mathrm{V} / m = 3 \times 10^{-6} \, \mathrm{cm}^2 \, \mathrm{g}^{-1}$. 
Given our resolution, we are not able to faithfully simulate a cross-section as large as $\sigma_\mathrm{V} / m = 3 \times 10^{-5} \mathrm{cm}^2 \mathrm{g}^{-1}$ because the mean free path becomes much smaller than the size of the kernel used for the self-interactions \citep[see Appendix~\ref{sec:mean_free_path} and][]{Fischer_2024b}.

We specify the strength of the cross-section in terms of the Knudsen number (Eq.~\ref{eq:knudsen_number}).
The collisionless case is indicated by $\mathrm{Kn} = \infty$ and the gaseous case by $\mathrm{Kn} = 0.0$. In Fig.~\ref{fig:dforce_vel} we can see that the dynamical friction is initially increasing (starting at the highest velocity) and decreases for later times of the simulation. At the beginning of the simulation, where the BH moves with a supersonic velocity, the dynamical friction is the smallest in the collisionless case. At later stages, it depends on the cross-section whether the self-interactions enhance or reduce the deceleration of the BH.
In Fig.\ref{fig:dforce_vel}, we plot the dynamical friction as a function of the velocity. This shows that the decelerating force reaches its maximum roughly at the sound speed, as expected.
Interestingly, the self-interacting run with an intermediate cross-section ($\mathrm{Kn} = 1.35 \times 10^{-1}$) shows the strongest dynamical friction, stronger than in the collisionless or gaseous regime. 
In principle, the figure is built analogously to Fig.~\ref{fig:ostriker}, but a direct comparison is limited by the fact that our simulation setup may not fulfil the assumptions of the Ostriker formula for the dynamical friction in the gaseous case (see Sect.~\ref{sec:analytic_estimates_dynfric}).
This is for example illustrated by Fig.~\ref{fig:cdm_vel_pos}, where we can see that the exact results depend on the choice of the setup, in particular on the box size.

Overall, we find that scatterings of the DM particles can alter the strength of the dynamical friction in a non-trivial way. Importantly, we have shown that the Knudsen number (Eq.~\ref{eq:knudsen_number}) provides a good measure for the medium being collisionless or collisional and its impact on the dynamical friction arising from the self-interactions.

\subsection{Properties of the wake} \label{sec:results_wake_properties}

\begin{figure}
    \centering
    \includegraphics[width=\columnwidth]{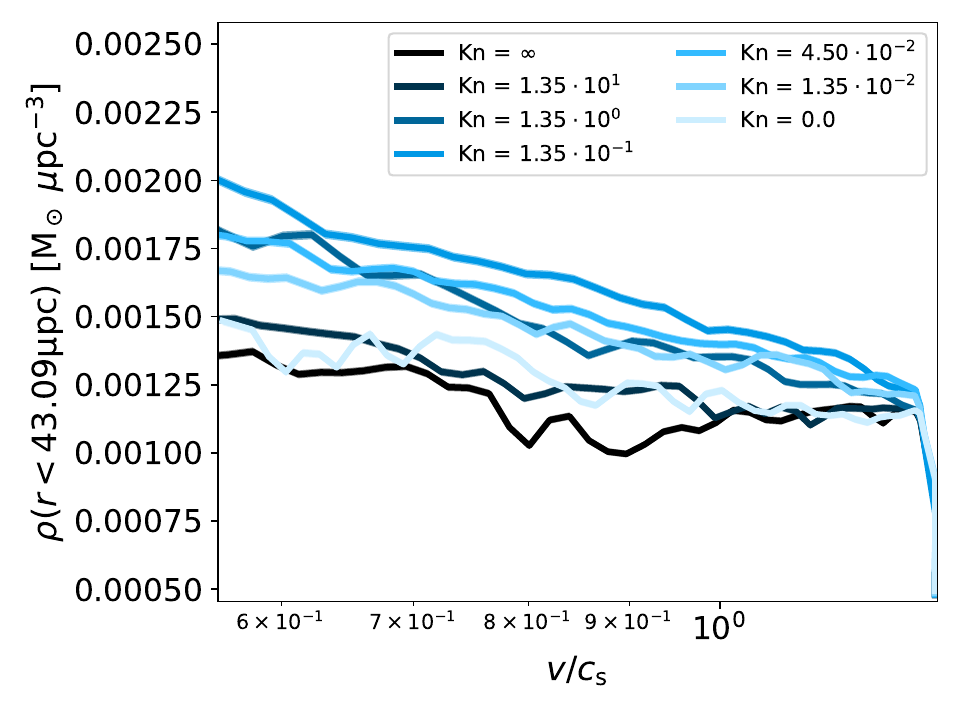}
    \caption{Density of the wake, i.e.\ the mean density about the point with the highest density, as a function of the velocity of the perturbing BH. The simulation results of the different cross-sections are indicated by their Knudsen number.}
    \label{fig:wake_density}
\end{figure}

\begin{figure}
    \centering
    \includegraphics[width=\columnwidth]{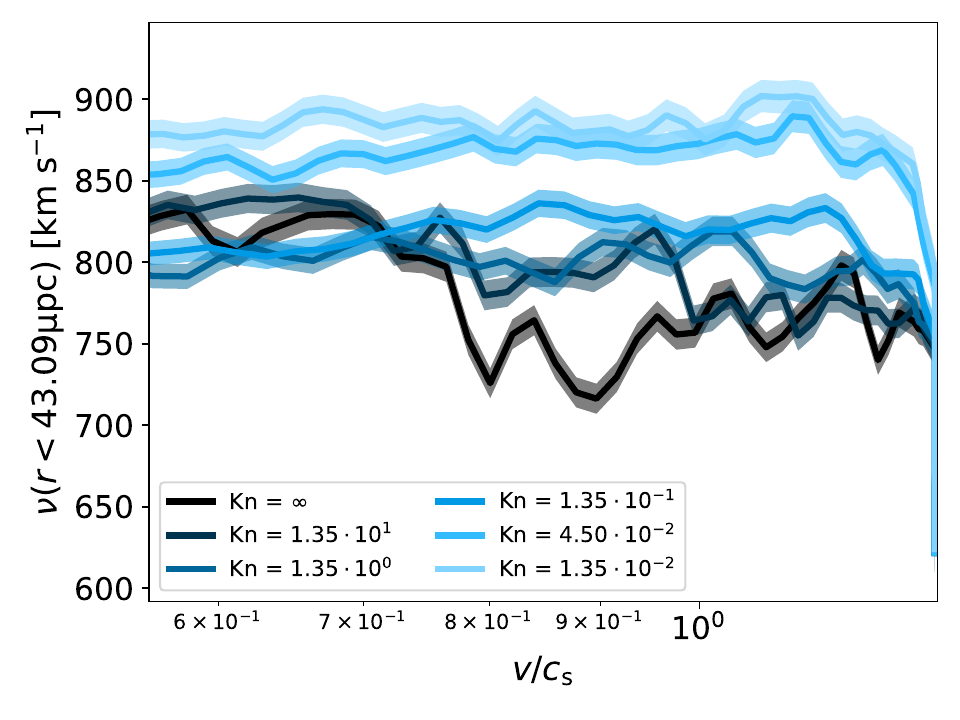}
    \caption{One-dimensional velocity dispersion of the wake, i.e.\ the velocity dispersion in the vicinity of the point with the highest density, as a function of the velocity of the perturber. We note that the velocity dispersion is computed after the bulk motion has been subtracted. The simulation results for different cross-sections are shown, indicated by the Knudsen number.}
    \label{fig:wake_vdispr}
\end{figure}

\begin{figure}
    \centering
    \includegraphics[width=\columnwidth]{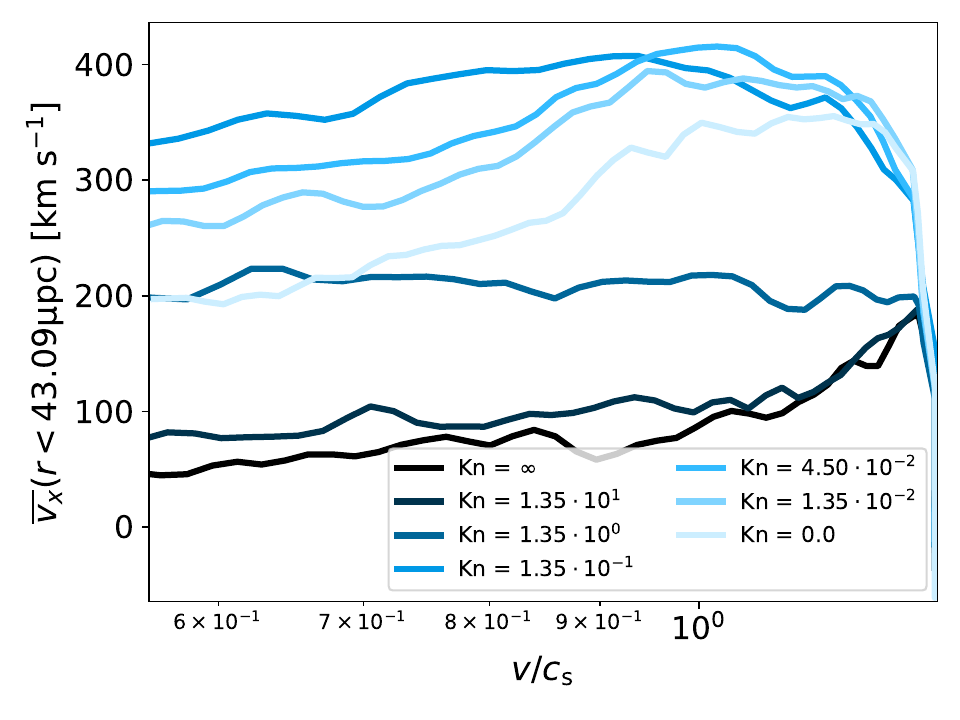}
    \caption{Mean velocity of the wake in the direction of the motion of the BH, i.e.\ the mean velocity in the vicinity of the point with the highest density. The results for different cross-sections are shown as a function of the velocity of the perturber. They are indicated by their corresponding Knudsen number.}
    \label{fig:wake_bulk_motion}
\end{figure}

\begin{figure}
    \centering
    \includegraphics[width=\columnwidth]{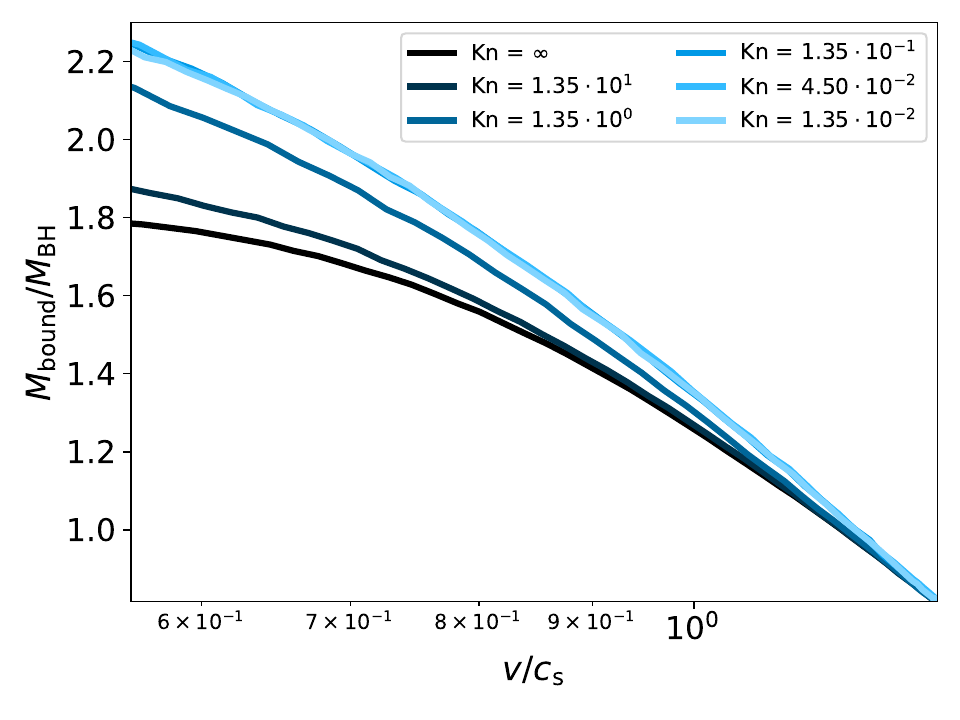}
    \caption{Dark matter mass gravitational bound to the perturbing BH as a function of the velocity of the perturber. We indicate the different cross-sections by their Knudsen number.}
    \label{fig:bound_mass}
\end{figure}

To provide more insight into the increased dynamical friction for mildly collisional media, we computed further quantities such as the density of the wake, its velocity dispersion, and bulk motion as well as the DM mass gravitational bound to the BH.

In Fig.~\ref{fig:wake_density}, we show the mean density within a sphere of radius $r=43.09\,\mu\mathrm{pc}$ about the point of the maximum DM density in the wake as a function of the velocity of the perturber.
The simulations showing a strong dynamical friction force also show a larger density compared to the ones with relatively weak dynamical friction.
This is not surprising, as a more massive wake can exert a stronger gravitational pull on the BH and thus faster decelerate it.
The presence of self-interactions allows for the growth of a larger wake density, as the scattering produces more low-velocity particles compared to the collisionless case. However, for sufficiently large self-interaction when entering the short-mean-free-path regime, the wake density starts to decrease with increasing cross-section.

Furthermore, we show the velocity dispersion in the density wake in Fig.~\ref{fig:wake_vdispr}.
We use the same volume as for the density computation.
The velocity dispersion was computed after subtracting the mean velocity of the particles.
The velocity dispersion in the wake increases compared to the ICs and is larger for stronger DM self-interactions.
Interestingly, the cross-section with $\mathrm{Kn} = 1.35 \times 10^{-1}$, which gives the strongest dynamical friction, shows a substantially lower velocity dispersion compared to the larger cross-sections.

The bulk motion of the density wake is displayed in Fig.~\ref{fig:wake_bulk_motion}.
The stronger the self-interactions the more efficiently the BH drags the DM along its trajectory, except for the largest cross-section for which the bulk velocity is again smaller.
This is in line with the strength of the dynamical friction force found in Fig.~\ref{fig:dforce_vel}, implying that when the DM is dragged along by the BH, it can more efficiently decelerate the BH.

Last, we computed the DM mass that is gravitationally bound to the BH.
To do so, we compared the relative velocity of a DM particle and the BH to the escape velocity,
\begin{equation}
    v_\mathrm{esc} = \sqrt{\frac{2\,\mathrm{G}\,m_\mathrm{BH}}{r}} \,,
\end{equation}
of the BH.
Here, the perturbing BH has a mass of $m_\mathrm{BH}$ and the gravitational constant is given by G.
The distance between the DM particle and the BH is denoted by $r$.
This may underestimate the bound mass to some extent as the DM bound to the BH increases the effective mass and thus may lead to further particles being gravitationally bound.

Our estimate for the gravitational bound mass is shown in Fig.~\ref{fig:bound_mass}.
It becomes visible that the bound mass depends on the cross-section.
In particular, strong self-interactions can effectively increase the bound mass.
This can be understood by the effect that the scattering has on the velocity distribution.
We note that the velocities in the wake do not follow a Maxwell--Boltzmann distribution, but the scattering evolves the distribution towards it.
This implies that the self-interactions create more low-velocity particles compared to the collisionless case.
In consequence, more DM particles are gravitationally bound to the BH.
This effectively enhances the dynamical friction acting on the perturber given the cross-section is small enough such that the behaviour known from the gaseous regime \citep{Ostriker_1999} does not become dominant.
Although the dynamical friction decreases with increasing cross-section for strong self-interactions, the gaseous regime nevertheless yields a stronger dynamical friction force in the supersonic regime compared to the collisionless case.

\subsection{Angular dependence}

The angular dependence of the self-interaction cross-section can in principle lead to qualitative differences in the evolution of astrophysical systems. This is for example well known for merging DM haloes \citep[e.g.][]{Fischer_2023b, Sabarish_2024} or the abundance of satellites \citep{Fischer_2022}.
However, for systems that are rather relaxed, the angular dependence plays a minor role only and the effect of self-interactions with different angular dependencies can be mapped onto each other using the viscosity cross-section $\sigma_\mathrm{V}$ \citep[Eq.~\ref{eq:sigma_v}, see also][]{Yang_2022D}.

To test whether this mapping also works for our setup to study dynamical friction, we ran simulations with a very anisotropic cross-section (using the fSIDM framework) and an isotropic cross-section.
The cross-sections are chosen to have the same value in terms of $\sigma_\mathrm{V} / m$ or $\sigma_\mathrm{T} / m$ to test the matching with the momentum transfer cross-section (see Eq.~\ref{eq:momentum_transfer_cross_section}) as well.

\begin{figure}
    \centering
    \includegraphics[width=\columnwidth]{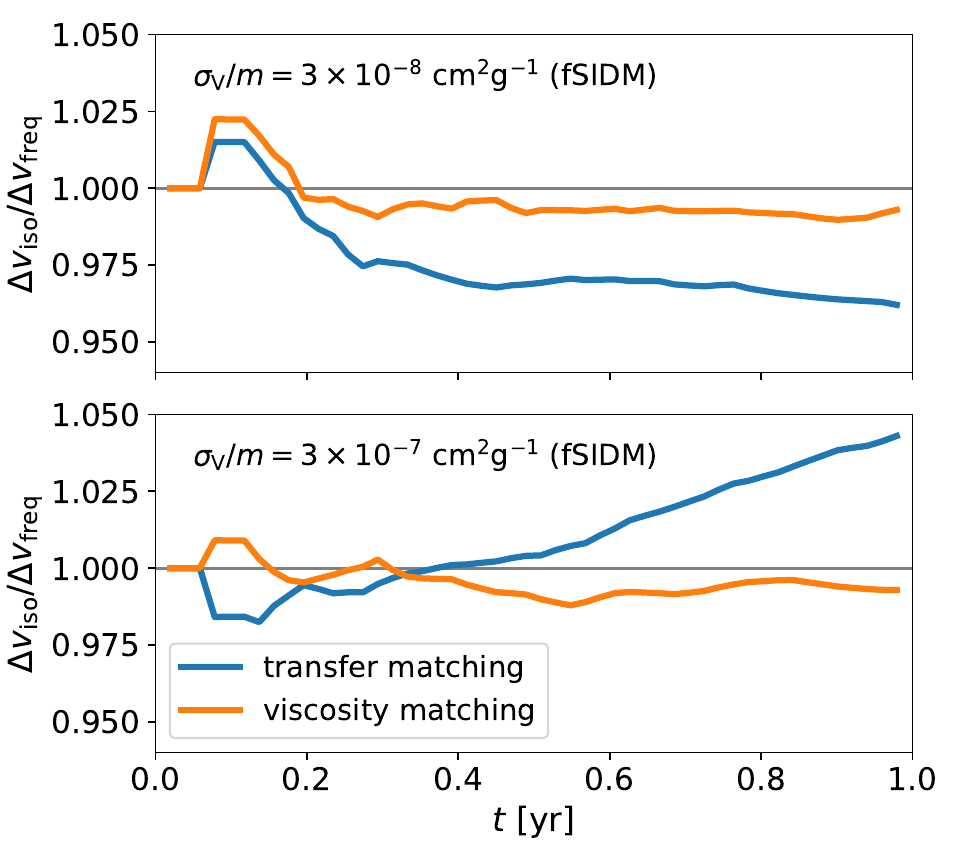}
    \caption{Ratio between the velocity change, $\Delta v = v(0) - v(t)$, due to dynamical friction of an fSIDM cross-section and the transfer (blue) / viscosity (orange) matched isotropic cross-sections. The upper panel is for an fSIDM viscosity cross-section of $\sigma_\mathrm{V}/m = 3 \times 10^{-8} \, \mathrm{cm}^2 \, \mathrm{g}^{-1}$ and the lower panel is for $\sigma_\mathrm{V}/m = 3 \times 10^{-7} \, \mathrm{cm}^2 \, \mathrm{g}^{-1}$.
    }
    \label{fig:matching}
\end{figure}

In Fig.~\ref{fig:matching}, we compare the velocity of the BH for the different angular dependencies.
We computed the ratio of the change in BH velocities for simulations with a matched isotropic cross-section to the fiducial fSIDM simulations.
The two panels in Fig.~\ref{fig:matching} display the ratio of the velocities as a function of time for a weaker fSIDM cross-section of $\sigma_\mathrm{V} / m = 3 \times 10^{-8} \, \mathrm{cm}^2 \, \mathrm{g}^{-1}$ (top panel) and a stronger one of $\sigma_\mathrm{V} / m = 3 \times 10^{-7} \, \mathrm{cm}^2 \, \mathrm{g}^{-1}$ (bottom panel).
For a perfect matching, we would expect the ratio to be unity. As we can see, the viscosity cross-section (Eq.~\ref{eq:sigma_v}) provides a better match than the transfer cross-section (Eq.~\ref{eq:momentum_transfer_cross_section}). This is the case for the weaker and stronger cross-section that we have simulated to test this.

\subsection{Velocity dependence}

The effective cross-section $\sigma_\mathrm{eff}$ \citep[introduced by][]{Yang_2022D} allows one to map cross-sections with different velocity dependencies onto each other.
It integrates the cross-section with its velocity dependence over the assumed Maxwell-Boltzmann distribution of the scattering velocities, applying a weighting factor of $v^5$.
The effective cross-section is given by
\begin{equation} \label{eq:sigma_eff2}
    \sigma_\mathrm{eff} = \frac{\langle v^5 \sigma_\mathrm{V}(v) \rangle}{\langle v^5 \rangle} \,.
\end{equation}
We note that this is very similar to the matching procedure proposed by \cite{Yang_2023S}; see also the work by \cite{Outmezguine_2023}.
It has been shown that this works fairly well for isolated haloes and, in turn, velocity-dependent cross-sections can be approximately described with constant cross-sections.
However, it is known that this does not work for systems that involve multiple velocity scales, such as merging galaxy clusters \citep{Sabarish_2024} or haloes containing substructure \citep{Fischer_2024a}.

\begin{figure}
    \centering
    \includegraphics[width=\columnwidth]{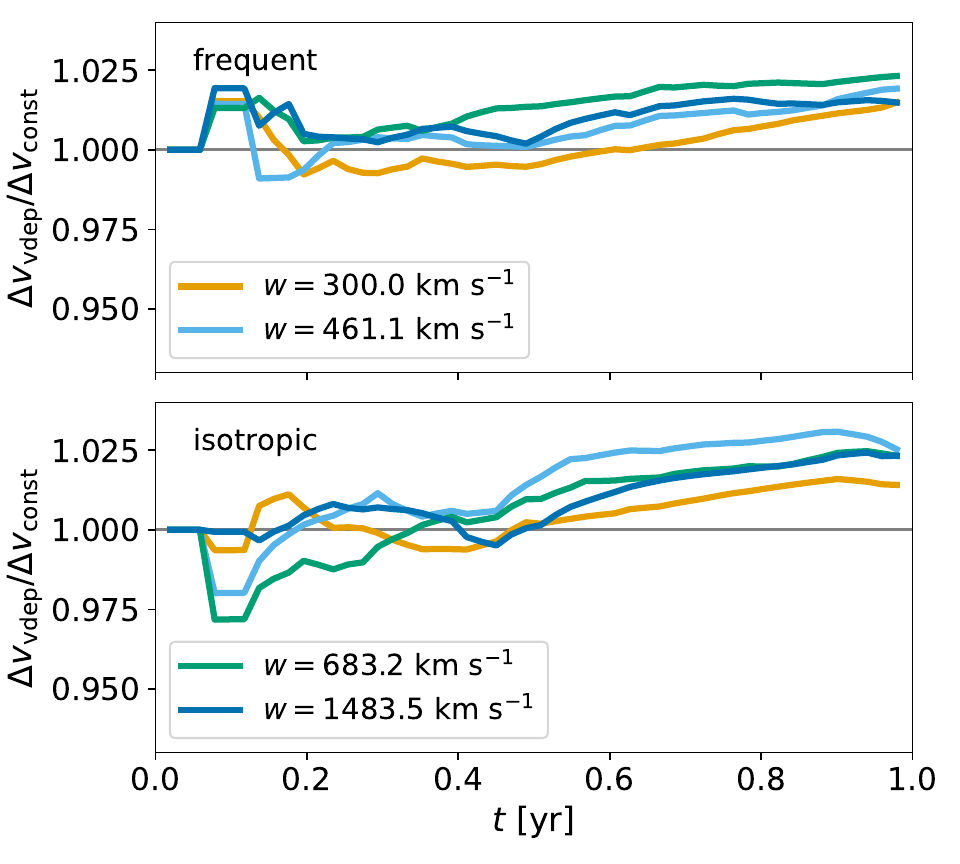}
    \caption{Ratio of the velocity change, $\Delta v = v(0) - v(t)$, due to dynamical friction acting on the BH for a velocity-dependent and a velocity-independent (constant) cross-section. All cross-sections were matched using the effective cross-section (Eq.~\ref{eq:sigma_eff}), i.e.\ all cross-sections correspond to $\sigma_\mathrm{eff} / m = 3 \times 10^{-8} \, \mathrm{cm}^2 \, \mathrm{g}^{-1}$. The upper panel displays runs with frequent self-interactions, and the lower panel is for isotropic scattering. 
    }
    \label{fig:vdep}
\end{figure}

Here, we investigate how well the matching works for our setup to study dynamical friction.
In Fig.~\ref{fig:vdep}, we show simulations involving different velocity dependencies specified in terms of $w$, but they are all matched using Eq.~\eqref{eq:sigma_eff}.
We have simulated very anisotropic cross-sections (top panel) and isotropic ones (bottom panel).
For times up to $t \approx 0.5 \, \mathrm{yr}$, the difference between the runs is fairly small and increases at later stages when the BH has slowed down.
We find that the matching of the velocity dependence using the effective cross-section does not work as well as for the angular dependence using the viscosity cross-section.
However, in our case where the velocity of the BH is falling below half its initial value over the course of the simulation, the deviation between the runs hardly exceeds 2.5\%.
This might be enough for a rough estimate, and separate simulations for velocity-dependent cases may only be needed when a higher accuracy is required.

\section{Discussion} \label{sec:discussion}

In this section, we discuss the limitations of our study, mainly the simulation setup. Moreover, we elaborate on the implications of our results and further directions to provide a better understanding of the dynamical friction in merging binary BHs.

Currently, there does not exist a theory for the non-equilibrium dynamics of SIDM to describe dynamical friction.
This is why we have to use simulations to understand the impact of a mildly collisional medium on the dynamical friction force.
Ideally, one would develop a more general formalism that contains both the Chandrasekhar and the Ostriker formula in the limits of a collisionless and a gaseous medium but also describes the intermediate regime as a function of the self-interaction cross-section. But even then one may still depend on simulations to calibrate such a formalism and make it applicable to problems such as merging binary BHs.
Moreover, we want to note that the Ostriker formula assumes the trajectory of the perturber to be a straight line. \cite{Kim_2007} improved on this and provided a generalisation to circular orbits.

For binary BHs, we find that the motion of the secondary BH through the DM spike would typically happen at Mach numbers larger than unity, in other words, in the supersonic regime.
Moreover, the Mach number does only very slightly depend on the radius, as the mass of the DM spike in the relevant radial range is typically much smaller than the mass of the primary BH.
Also, the spike index does not play a significant role in the motion of the secondary BH being supersonic.
But more importantly, a non-zero eccentricity could lead to a subsonic phase about the apocentre of the orbit of the secondary BH.
This implies that the supersonic regime where the dynamical friction is enhanced by self-interactions plays an important role and would lead to a faster decay of the orbit compared to a collisionless medium with the same density.

To study the dynamical friction with self-interactions being present, simulations are required.
We used an idealised setup with a periodic box hosting initially a constant density.
We have to note that this setup comes with drawbacks that limit the conclusion we can draw.
In particular, it is difficult to create a stable setup where the box size is small enough such that the velocity dispersion of the DM counteracts the gravitational collapse.
This effectively limits the maximal box size that can be studied, which also has an impact on the dynamical friction.
This is because the box size is related to the maximum impact parameter $r_\mathrm{max}$ (see Eq.~\ref{eq:chandrasekhar}).
Moreover, the setup does not always fulfil the assumptions of analytic formulas discussed in Sect.~\ref{sec:analytic_estimates_dynfric}.
Also, we cannot exclude that perturbations arising from the supersonic phase still affect the trajectory of the BH in the subsonic phase.
Nevertheless, we are still able to make qualitative statements and gain new insights into the dynamical friction.
Especially for this purpose, we tuned the setup to give a relatively larger impact on the BH velocity and make it easier to obtain qualitative differences being much larger than the numerical errors.

To obtain accurate quantitative predictions on the strength of the dynamical friction in BH mergers, one would need to simulate a setup much closer to the real system.
\cite{Kavanagh_2020, Kavanagh_2024} studied intermediate mass-ratio inspirals resolving the DM spike around the primary BH and measured the deceleration due to the dynamical friction of the secondary.
In principle, such a setup could be used for studies of SIDM as well.
Potentially this does not only allow us to calibrate the Chandrasekhar formula, but also the Ostrikers formula for the gaseous regime and with this make it applicable for merging BHs.
Ideally, one would have a more general formalism for arbitrary cross-sections that can be calibrated as well.

In our simulations, we find a regime of intermediate cross-sections where the dynamical friction is stronger than in the collisionless case or the gaseous case.
Given the drawback our setup suffers from, it would be great to study this in a setup closer to the physical systems as described above.
Simulations of BHs moving through a DM spike could potentially tell whether the intermediate regime of enhanced dynamical friction is only due to our setup or would also occur in real BH mergers.

The prediction of the exact strength of the dynamical friction force and its impact on the GW signal is difficult.
We find that not only does the DM density matter but its collisionless or collisional nature, too.
Moreover, the latter plays a crucial role in the spike index and the DM density \citep{Shapiro_2014}.
But it must also be noted that the energy that is transferred due to dynamical friction from the secondary BH into the DM spike affects the DM spike as well.
\cite{AlonsoAlvarez_2024} argues that only SIDM can effectively transport this heat outwards and replenish the DM spike.
In contrast to collisionless matter, this effectively allows a SIDM spike to make the BH orbit decay and solve the final parsec problem.
Overall, it remains challenging to predict the exact GW signal, especially beyond the Newtonian regime.

\section{Conclusions} \label{sec:conclusion}

    In this paper, we have reviewed analytic estimates for the dynamical friction force in the collisionless and gaseous regime, and discussed the two regimes in the context of binary BH mergers. Furthermore, we have performed idealised simulations to study the dynamical friction as a function of the self-interaction cross-section. In addition, we also studied how well cross-sections with different angular and velocity dependencies can be matched onto each other.
    Our main results can be summarised as follows:
    \begin{enumerate}
      \item DM self-interactions can enhance but also lower the dynamical friction force, depending on the velocity of the perturber relative to the local velocity dispersion of the DM.
      \item For merging binary BHs, we expect the perturber to be in the supersonic regime all the time when being in a circular orbit and partially when following an eccentric orbit. In this regime, self-interactions enhance the dynamical friction force.
      \item The Knudsen number, Kn, is a good proxy for how much the self-interactions impact the dynamical friction, in other words, how close the system is to the collisionless or gaseous regime.
      \item When the SIDM cross-section is large enough, $\mathrm{Kn} \lesssim 2$, we expect the dynamical friction in the supersonic regime to be significantly larger than in the collisionless case.
      We note that for typical binary BHs such a cross-section could be many orders of magnitude smaller than current upper bounds on the self-interaction cross-section \citep{Adhikari_2022}.
      \item This implies that observed BH mergers, when being consistent with collisionless matter, can potentially imply extremely tight upper bounds on the SIDM cross-section.
      \item We find that the viscosity cross-section provides a good matching procedure between isotropic and very anisotropic cross-sections when computing the effect of self-interactions on the dynamical friction.
      \item The matching of different velocity dependencies onto each other works less well than the angular matching. However, depending on the application, the deviation could still be within an acceptable range.
   \end{enumerate}

   The results we obtained are solely based on simulations of a constant density perturbed by a BH in a periodic box.
   This allowed us to gain insight into the qualitative behaviour but does not allow for accurately quantifying how strong the effect of self-interactions is on the dynamical friction force.
   In consequence, the next step would be to simulate a SIDM spike with an orbiting BH, similar to the work done by \cite{Kavanagh_2020, Kavanagh_2024}.
   Such a setup would allow us to precisely determine the strength of the dynamical friction force acting onto BHs orbiting in DM spikes when self-interactions are present.
   This is the subject of a forthcoming paper (Sabarish et al., in prep).


\begin{acknowledgements}
The authors thank the organisers of the Pollica 2023 SIDM Workshop, where part of this work was done. 
In addition, the authors thank all participants of the Darkium SIDM Journal Club for discussions.
MSF and LS thank Gonzalo Alonso-Álvarez, Niklas Becker, Andrew Benson, Akaxia Cruz, Xiaolong Du, Frederick Groth, Charlie Mace, Annika H.\ G.\ Peter, Rainer Spurzem, Sabarish Venkataramani, Daneng Yang, Shengqi Yang, Hai-Bo Yu, Xingyu Zhang, and Zhichao Carton Zeng as well as the members of the CAST group at the University Observatory Munich for helpful discussions.
They are also grateful to Bradley Kavanagh for helpful comments on the draft.
This work is funded by the Deutsche Forschungsgemeinschaft (DFG, German Research Foundation) under Germany’s Excellence Strategy -- EXC-2094 `Origins' -- 390783311.
Software: NumPy \citep{NumPy}, Matplotlib \citep{Matplotlib}.
\end{acknowledgements}

%
%

\bibliographystyle{aa}
\bibliography{bib.bib}

\begin{appendix}
\section{Unresolved mean free path} \label{sec:mean_free_path}
\begin{figure}[!ht]
    \centering
    \includegraphics[width=\columnwidth]{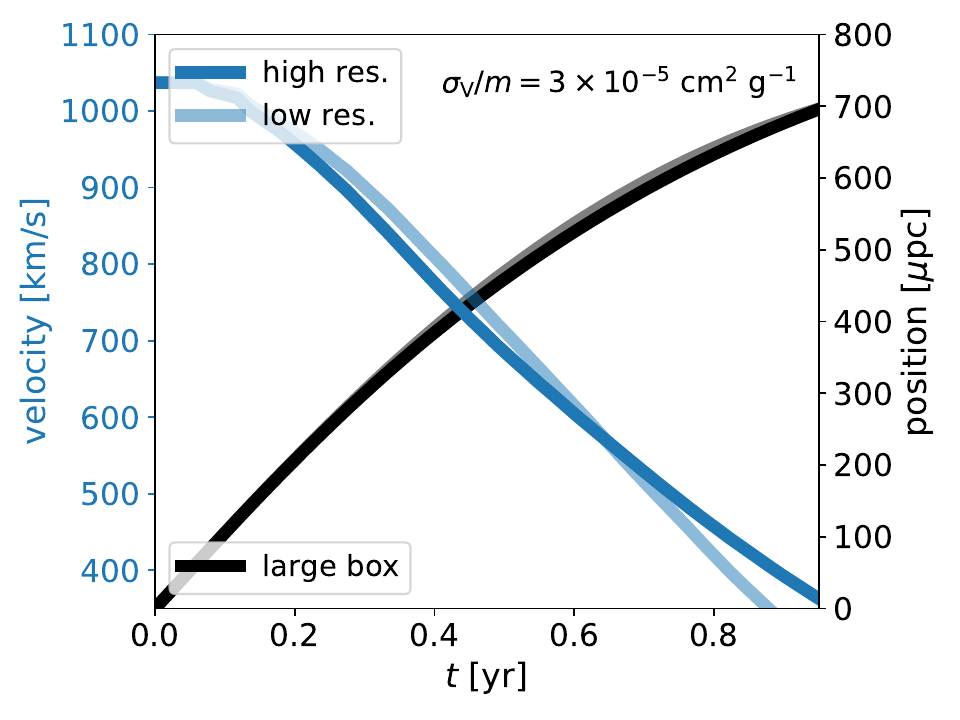}
    \caption{Position and velocity of the BH in our test setup for a simulation with a fSIDM cross-section of $\sigma_\mathrm{V} / m = 3 \times 10^{-5} \, \mathrm{cm}^2 \, \mathrm{g}^{-1}$. The high resolution corresponds to $N=5 \times 10^6$ particles and the low resolution to $N = 1 \times 10^6$ particles. The larger box size of $l_\mathrm{box} = 829.5 \, \mu \mathrm{pc}$ was used.}
    \label{fig:f1e-5}
\end{figure}

In this appendix, we show the results of two simulations with a larger cross-section than the ones we have shown in the main text.
These additional simulations are for fSIDM and employ a cross-section of $\sigma_\mathrm{V} / m = 3 \times 10^{-5} \, \mathrm{cm}^2 \, \mathrm{g}^{-1}$. They differ in resolution, we use $N=1 \times 10^6$ and $N = 5 \times 10^6$ particles.
In Fig.~\ref{fig:f1e-5}, we can see, in contrast to the previous simulations, that the results for the two resolutions differ significantly from each other.

With increasing cross-section, the mean free path is shrinking and may fall below the kernel size used for SIDM \citep[see][]{Fischer_2024b}. The ratio between the kernel size, $h$, and the mean free path $l$, is fairly large for these simulations, $h/l \approx 172$ for the lower resolution run and $h/l \approx 100$ for the higher resolution. Thus, it is not surprising that the simulations deviate from each other.
The value for $h/l$ at which the simulations start to become inaccurate may in general depend on the problem at hand. The values of our simulations here may not be indicative of other setups.

\end{appendix}

\end{document}